\documentclass[12pt,a4paper]{article}
\usepackage{amsmath}

\title{Five Specific Cases of the Simple Equations Method (SEsM)}

\author{Zlatinka I. Dimitrova}

\date{Institute of Mechanics, Bulgarian Academy of Sciences, 1113 Sofia, Bulgaria\\  e-mail: zdim@imbm.bas.bg}

\begin{document}
\maketitle

\begin{abstract}
We discuss the Simple Equations Method (SEsM) for obtaining exact solutions of 
nonlinear partial differential equations. We show that 
the  Jacobi Elliptic Function Expansion Method, F-Expansion method, Modified Simple Equation method, Trial Function Method, General Projective Riccati Equations Method 
and the First Integral Method are specific cases of SEsM.
\end{abstract}
\section{Introduction}
Many complex systems are nonlinear  \cite{a1}-
\cite{a4}. Usually, the effects connected to nonlinearity   are studied  by the
methods of  the time  series analysis or are modeled  by nonlinear differential or 
difference equations \cite{n1} - \cite{t10}. The methodology
for obtaining exact solutions of nonlinear differential equation was in
development since many years. At the beginning one tried
 to remove the nonlinearity of the solved equation  
by means of appropriate transformation,e.g., the Hopf-Cole 
transformation \cite{hopf}, \cite{cole}  transforms the nonlinear 
Burgers  equation to the linear heat equation. Such attempts leaded to the
development of    the Method of Inverse Scattering 
Transform \cite{ablowitz} - \cite{gardner} and  the  method of Hirota  \cite{hirota}, 
\cite{hirota1} .   Kudryashov and then Kudryashov and Loguinova developed   the 
Method of Simplest Equation (MSE)
\cite{k05},\cite{kl08}  based on determination of singularity order $n$ of the solved NPDE and searching of
particular solution of this equation as series containing powers of solutions
of a simpler equation called simplest equation. 
Below we discuss   some aspects of a methodology for obtaining exact and approximate
solutions of nonlinear partial differential equations called  Simple 
Equations Method (SEsM) \cite{se1} - \cite{se8}. 
The development of this methodology started  in the 1990's 
\cite{mv1} - \cite{mv5}. A specific case of the methodology  based on use of one simple equation was used  in 2009
\cite{1}, \cite{2}. In  2010  the ordinary differential 
equation of Bernoulli as simplest equation \cite{v10} was used as a simple equations  the corresponding version of  SEsM  wa called Modified Method of Simplest Equation (was) and it was applied to ecology
and population dynamics \cite{vd10}.  The MMSE  
\cite{vdk}, \cite{v11} is based on determination of the kind
of the simplest equation and truncation of the series of solutions of the 
simplest equation by means 
of application of a balance equation and it is equivalent of the MSE mentioned above. 
\par 
We used MMSE for various applications till 2018
\cite{v11a} - \cite{vdv17}. An interesting paper from this period is \cite{vdv15}
where  we have extended the  MMSE to simplest equations of the class
\begin{equation}\label{sp_func}
\left (\frac{d^k g}{d\xi^k} \right)^l = \sum \limits_{j=0}^{m} d_j g^j
\end{equation} 
where $k=1,\dots$, $l =1,\dots$, and $m$ and $d_j$ are parameters. The solution 
of the last equation contains as specific cases, e.g.,: trigonometric 
functions;  hyperbolic functions;  elliptic functions of Jacobi;  
elliptic function of Weierstrass, etc. Recently, Vitanov extend the capacity of 
the methodology by inclusion of the possibility of use of more than one simplest 
equation. This modification  is called   SEsM - Simple Equations Method as 
the used simple equations are more simple than the solved nonlinear 
partial differential equation but these simple equations in fact can be quite 
complicated.   We note that a variant of SEsM based on two simple equations was 
applied in \cite{vd18} and the first description of the methodology was made in 
\cite{se1} and then in \cite{se2} - \cite{se8}. For more applications of specific cases of the  methodology see \cite{n17} - \cite{vnew24}. 
\par
 The goal of this article is to show that several frequently used methods for obtaining exact solutions of nonlinear partial differential equations are  specific 
cases of SEsM. The organization of the text is as follows.
 We  discuss the SEsM  
in Sect 2. In Sect. 3 we show that Jacobi Elliptic Function Expansion Method and F-Expansion method are specific cases of SEsM.  In Sect. 4 we show that the Modified Simple Equation Method Method is specific case of SEsM.
In Sect. 5 we show that the Trial Function Method is a specific case of SEsM.
In Sect. 6 we show that the General Projective Riccati Equations Method is a
specific case of SEsM.
In Sect. 7 we show that the First Integral Method is specific case of SEsM 
Several concluding remarks are summarized in the Sect.8 of the article.
\section{The Simple Equations Method (SEsM)}
The methodology of SEsM has 4 steps. They are  as follows. 
Let us consider a  a system of nonlinear partial differential equations
\begin{equation}\label{ses1}
{\cal W}_i [u_1(x,\dots,t), \dots, u_n(x,\dots,t) ] = 0, i=1,\dots,n. 
\end{equation}
Above, ${\cal W}_i [u_1(x,\dots,t), \dots, u_n(x,\dots,t) ]$ depends on the functions $u_1(x,\dots,t), \dots, u_n(x,\dots,t) $ and some of their derivatives ($u_i$ can be a function of more than 1
spatial coordinates). Step 1 of SEsM is connected to the  transformations
\begin{equation}\label{ses2}
u_i(x, . . . , t) = T_i[F_i(x, \dots, t), G_i(x,\dots,t), \dots]
\end{equation}
where $T_i(F_i,G_i, \dots)$ is some function of another functions $ F_i,G_i,\dots$. In general $ F_i(x, \dots , t)$, $G_i(x,\dots,t)$, $\dots$  are  functions of several spatial variables
as well as of the time. The transformations has the goal  to remove 
the nonlinearity of the solved differential equations or to transform this nonlinearity to
more treatable kind of nonlinearity or the transformation may even remove the nonlinearity. Several example for the transformations $T(F,G,\dots)$  in the case of one solved equation are
\begin{description}
\item[Specific case 1:] the Painleve expansion,
\item[Specific case 2:] $u(x,t)=\frac{\sum \limits_{i=0}^I a_i [F(x,t)]^i}{\sum \limits_{j=0}^J b_j [G(x,t)]^j}$,
\item[Specific case 4:] $u(x, t) = 4 \tan^{-1}[F(x, t)]$ for the case of the sine - Gordon equation.
\end{description}
In some  cases one may skip this step  but in numerous other
cases the step is necessary for obtaining a solution of the studied nonlinear PDE. The application of (\ref{ses2}) to
(\ref{ses1}) leads to a nonlinear PDEs  for the functions $F_i,G_i,\dots$.
\par 
 Step 2. of SEsM follows. In this step, the functions $F_i(x, . . . , t)$, $G_i(x,\dots,t)$, $\dots$ are 
represented as a function of other functions $f_{i1}, . . . , f_{iN}$,
$g_{i1},\dots,g_{iM}$, $\dots$,   which are connected to solutions
of some differential equations (these equations can be partial or ordinary 
differential equations) that are more
simple than Eq.(2). We note that the possible values of N and M are N = 1, 2, . . 
., M=1,2,... (there may be infinite number of
functions f too). The forms of the functions $F_i(f_1,\dots,f_N)$, $G_i(g_1,
\dots,g_M)$, $\dots$ can be 
different. One example for the function $F$ in  the case of one solved equation is
\begin{equation}\label{hir} 
		F = \alpha + \sum \limits_{i_1=1}^N \beta_{i_1} f_{i_1} + \sum \limits_{i_1=1}^N  \sum \limits_{i_2=1}^N 
		\gamma_{i_1,i_2} f_{i_1} f_{i_2} + \sum \limits_{i_1=1}^N \dots \sum \limits_{i_N=1}^N \sigma_{i_1,\dots,i_N} f_{i_1} \dots f_{i_N}.
\end{equation} 
Here, $\alpha,\beta_{i_1}, \gamma_{i_1,i_2}, \sigma_{i_1,\dots,i_N}\dots  $ are parameters.  $F(f_1,\dots,f_N)$ can have also different form. 
We note that the  relationship (\ref{hir}) contains the 
		relationship used by Hirota \cite{hirota} as specific case
\par 
In Step 3. of SEsM, we have to represent  the functions used in $F_i, G_i, \dots$. This means that we choose the PDEs which are solved by the functions $f_{i1},\dots,f_{iN}$, $g_{i1}, \dots, g_{iM}$.  These equations are more
simple than the solved nonlinear partial differential equation. 
One may use solutions of the simple partial differential equations 
for $f_{i1},\dots,f_{iN}$, $g_{i1}, \dots, g_{iM}$
if such solutions are available, or may
transform the more simple partial differential equations by means of appropriate ans{\"a}tze. Then the solved differential equations for
$f_{i1}$, $\dots$, $f_{iN}$, $g_{i1}, \dots, g_{iM}$, $\dots$ can be reduced to differential equations $E_l$, containing derivatives of one or several functions
\begin{equation}\label{i1}
E_l \left[ a(\xi), a_{\xi},a_{\xi \xi},\dots, b(\zeta), b_\zeta, b_{\zeta \zeta}, \dots \right] = 0; \ \
		l=1,\dots,N+M+\dots.
	\end{equation} 
Next, we assume that the functions $a(\xi)$, $b(\zeta)$, etc.,  are  functions of 
	other functions, such as, $v(\xi)$, $w(\zeta)$, etc., e.g,
	\begin{equation}\label{i1x}
	a(\xi) = A[v(\xi)]; \ \ b(\zeta) = B[w(\zeta)]; \dots.
	\end{equation} 
	Note that SEsM does not prescribe the forms of the functions $A$ , $B$, $\dots$. 
	Often one uses a finite-series relationship, e.g, 
	\begin{equation}\label{i2}
	a(\xi) = \sum_{\mu_1=-\nu_1}^{\nu_2} q_{\mu_1} [v (\xi)]^{\mu_1}; \ \ \ 
	b(\zeta) = \sum_{\mu_2=-\nu_3}^{\nu_4} r_{\mu_2} [w (\zeta)]^{\mu_2}, \dots .
	\end{equation}
	where $q_{\mu_1}$, $r_{\mu_2}$, $\dots$ are parameters.
	However, other kinds of relationships may also be used.
\par  Finally, at this step of SEsM, we choose the simple differential equations
which are solved by the functions  $v(\xi)$, $w(\zeta)$, $\dots$. 
Then,  we apply  the steps 1.) - 3.) to Eqs.(2) and usually this transforms  the left-hand side of these equations to a function which is a sum of 
terms where each term contains some
function multiplied by a coefficient. This coefficient contains some of the 
parameters of the solved equations and
some of the parameters of the solution. In the most cases a balance procedure must be applied in order to ensure
that the above-mentioned relationships for the coefficients contain more than one term. This balance procedure may lead to one or more
additional relationships among the parameters of the solved equation and parameters of the solution. These additional
relationships are called balance equations.
\par 
Finally at Step 4. of SEsM We can obtain a nontrivial solution of Eq. (2) if all coefficients mentioned above in the text are set to 0. This leads usually to a system of nonlinear algebraic equations for the coefficients of the solved nonlinear
PDE and for the coefficients of the solution. Any nontrivial solution of this algebraic system leads to a solution
the studied nonlinear partial differential equation. Usually the above system of algebraic equations contains
many equations that have to be solved with the help of a computer algebra system.
\section{Jacobi Elliptic Function Expansion Method and F-expansion Method as Specific Cases of  SEsM}
The organization of this Section is as follows.
\begin{enumerate} 
\item  
We prove first that the Jacobi Elliptic Function Expansion Method (JEFEM)
in its classic from is specific case of SEsM. 
\item  
We describe General Jacobi 
Elliptic Function Expansion Method (GJEFEM) and prove that it is specific case 
of SEsM. 
\item  
We list several methods used in the literature which are
specific cases of GJEFEM.
\end{enumerate} 
\par 
The classic from of JEFEM is as follows \cite{liu01}.
One considers nonlinear partial differential equation for $u(x,t)$ in the
form
\begin{equation}\label{ell1}
N(u,u_x,u_t,u_{xx},u_{xt},u_{tt},\dots)=0
\end{equation}
and searches for traveling wave solutions in the form
\begin{equation}\label{ell2}
u = u (\xi): \ \ \xi = k(x-ct).
\end{equation}
Above, $k$ and $c$ are parameters. $u(\xi)$ is searched in the form of series of
the Jacobi elliptic function ${\rm sn}(\xi,m)$ where $m$ is the modulus 
of the function ${\rm sn}$,
\begin{equation}\label{ell3}
u(\xi) = \sum \limits_{j=0}^n a_j {\rm sn}(j,m)^j
\end{equation}
This is a generalization of the $\tanh$-method because for 
$m=1$ ${\rm sn}(\xi,1) = \tanh(\xi)$.
The substitution of (\ref{ell2}) and (\ref{ell3}) in (\ref{ell1}) can lead
to a system of nonlinear algebraic equations and any nontrivial solution
of this system leads to an exact traveling wave solution of the solved equation
(\ref{ell1}).
\par 
{\bf Proposition}\\
{\em 
The Jacobi Elliptic Function Expansion  Method  (JEFEM) is a specific case of SEsM for the case when there is no transformation of the nonlinearity of the equation (Step 1 of SEsM is skipped); Function $F$ at Step 2 of SEsM has specific form -  (\ref{ell3})); just one simple equation is used and this simple equation is the differential equation for the elliptic function ${\rm sn}$.}
\par 
{\bf Proof}\\
\begin{enumerate}
	\item   
We start from SEsM, impose restrictions on it and reduce SEsM to JEFEM. 
\item  
In Step 1. of SEsM we do not transform the nonlinearity of
the solved equation (we just skip this step).
Additional restriction is that we search for traveling wave solution of the
solved equation (\ref{ell1}).
\item  
In Step 2. of SEsM we use one of the possible forms of the function $F$,
namely, the form (\ref{ell3}).
\item 
In Step 3. of SEsM  we use the function from (\ref{ell3}) as the JEFEM is directly connected to the solution of the used simple equation which is the
equation for the Jacobi elliptic function ${\rm sn}$. We note that the use of only one simple equation is a further restriction on SEsM. By means of all
restrictions above we reduce SEsM to JEFEM. Thus JEFEM  is specific case of SEsM
\end{enumerate}  
\par
Now let us formulate General Jacobi Elliptic Function Expansion  Method
(GJEFEM). In this method we solve in general a system of $N$ nonlinear partial
differential equations and search for traveling wave solutions based on different
coordinates $\xi_i = \alpha_i x - \beta_i t$, $i=1,2,\dots,N$. The solution is
searched as function
\begin{equation}\label{ell4}
u_i(\xi_1,\dots,\xi_n) = U_i[f_1(\xi_1),\dots,f_N(\xi_N)]
\end{equation} 
of the functions $f_1,\dots, f_N$ and each of these functions is a solution
of a differential equations for the Jacobi elliptic functions
\begin{equation}\label{ell5}
\left(\frac{df_i}{d \xi_i} \right)^2 = a_i f_i^4 + b_i f_i^2 + c_i .
\end{equation}
Next, we show that the GJEFEM is a specific case of SEsM.
\par 
{\bf Proposition}\\
{\em The General Jacobi Elliptic Function Expansion  Method  (GJEFEM) is specific 
cases of SEsM for the case when there is no transformation of the nonlinearity of 
the equation (Step 1 of SEsM is skipped); Functions $u_i$ at Step 2 of SEsM have 
specific form -  (\ref{ell4})); and the simple equations are of the kind of the 
differential equation for the Jacobi elliptic functions.} 
\par 
{\bf Proof}\\
We start from SEsM, impose restrictions on it and reduce SEsM to JEFEM. 
\begin{enumerate}
	\item 
In Step 
1. of SEsM we do not transform the nonlinearity of
the solved equation (we just skip this step).
Additional restriction is that we search for traveling wave solution of the
solved equation (\ref{ell1}).
\item  
In Step 2. of SEsM we use a possible form of the functions $u_i$ - (\ref{ell4}).
In  Step 3.  of SEsM the functions from (\ref{ell4}) in the JEFEM 
are directly connected to the solution of the used simple equations which 
are of the kind of the differential equation for the Jacobi elliptic functions.
This is additional restriction on SEsM. 
\item  
By means of all
restrictions above we reduce SEsM to GJEFEM. Thus GJEFEM  is specific case of 
SEsM.
\end{enumerate} 
\par 
Let us now list several specific cases of GJEFEM. 
\begin{enumerate} 
\item 
JEFEM is specific case
of GJEFEM for the case of just one solved nonlinear partial differential equation 
and when the simple equation is the equation for the Jacobi elliptic function $
{\rm sn}$ and in addition the function $U$ is a power series of the 
function ${\rm sn}$. 
\item  
Parks et al. \cite{p02} and Fu et al. \cite{fu} use expansions based on the elliptic functions ${\rm cn}$, ${\rm dn}$ and 
${\rm cs}$. This is specific case of 
GJEFEM when one simple equation is used and this simple equation is of the 
kind of (\ref{ell5}). 
\item  
Fan and Zhang \cite{fz} present interesting application
which is extension of JEFEM for the case of two functions $u_{1,2}$ and
single simple equation and by means of this extension they obtain solutions
of the coupled Schr{\"o}dinger - KdV system and of two-dimensional Davey –
Stewartson equation. This extension of JEFEM is specific case of GJEFEM
when two functions $u_{1,2}$ are used with the same argument and when the
simple equation is the differential equation for the elliptic function ${\rm sn}
$. 
\item  
Another specific case of GJEFEM was applied by Yan \cite{yan} who 
treated  a (2 + 1)-dimensional integrable Davey - Stewartson - type equation
for the case of 2 spatial coordinates and travelling wave solutions. We note
that SEsM allows treating equations with more that one spatial coordinate
and the travelling waves can travel with different velocities which is more 
general case than the case discussed by Yan where we have a single traveling wave
despite the two spatial coordinates presented. Yan uses the following form
of the function $u_i$, $i=1,2,3$
\begin{equation}\label{ell6}
u_i (\xi) = a_{i0} + \sum \limits_{j=1}^n  f_k^{j-1}(\xi)[a_{ij}f_k(\xi) + 
b_{ij} g_k(\xi)]
\end{equation}
where $f_k$ and $g_k$, $k=1,\dots,12$ are Jacobi elliptic functions (i.e. are
functions which satisfy the simple equation of kind (\ref{ell5})). (\ref{ell6})
is specific form of the function $U_i$ from GJEFEM and the simple equations
are equations for Jacobi elliptic functions as in GJEFEM.
\item 
Another specific case of GJEFEM is used in \cite{pan}. The simple equations used there are for Jacobi elliptic functions and the specific case of the
used single function $U$ is
\begin{equation}\label{ell7}
U = a_0 + \sum \limits_{i=1}^N {\rm sn}^{-1}(\xi, m)[a_i {\rm sn}(\xi,m) + b_i
{\rm cn}(\xi,m)] 
\end{equation}
\item 
Liu and Fan \cite{l05} apply specific case of GJEFEM for the case of two 
spatial coordinates and time. These three variables are combined to produces a 
single traveling wave coordinate which allows the use of single variable
simple equations. 
\item 
Wang et al. \cite{w05} use also specific case of GJEFEM
for the case of two spatial variables and time and combine all these variables in
a single traveling wave variable. The new point in this article is the specific 
form of the functions $U_i$
\begin{equation}\label{ell8}
U_i = a_{i0} + \sum \limits_{j=1}^{m_1} \left[ 
a_{ij} \frac{{\rm sn}^j(\xi,m)}{(\mu {\rm sn}(\xi,m)+1)^j} + b_{ij} \frac{{\rm sn}^{j-1}(\xi,m) {\rm cn}(\xi,m)}{(\mu {\rm sn}(\xi,m)+1)^j} \right]
\end{equation}
\item  
Ye at al. \cite{yu05} extend (\ref{ell8}) and use the
following specific case for the functions $U_i$
\begin{equation}\label{ell9}
U_i = a_{i0} + \sum \limits_{j=1}^{m_1} \left[ 
\frac{{a_{i,2j-1}\rm sn}^j(\xi,m)}{(\mu {\rm sn}(\xi,m)+ \mu_2 {\rm cn}(\xi,m)+1)^j} +  \frac{a_{i,2j}{\rm sn}^{j-1}(\xi,m) {\rm cn}(\xi,m)}{(\mu {\rm sn}(\xi,m)+\mu_2 {\rm cn}(\xi,m)+1)^j} \right]
\end{equation}
\item 
Other variants for $U_i$ are proposed by Wang et al. \cite{wx05}, Chen and Wang 
\cite{c05}, L{\"u} \cite{lu}, Abdou and Elhanbaly \cite{a07}, El-Sabbagh and Ali \cite{s08}, 
\cite{a11}.
\item 
Another specific case of GJEFEM is the F-expansion method which has the same
ideology as JEFEM but only the form of the simple equations for the Jacobi
elliptic functions are not specified. In the different variants of the F-
expansion method one uses different specific cases for the functions $U_i$
from GJEFEM \cite{z1}, \cite{wx2}, \cite{wx3}, \cite{ren}.
\end{enumerate} 

\section{Modified Simple Equation Method as Specific Case of SEsM}
The Modified Simple Equation Method is as follows 
\cite{jawad10}. One considers the nonlinear partial differential equation which can be reduced to an
ordinary partial differential equation for the function $u(z)$
\begin{equation}\label{se1}
P(u,u_z,u_{zz},u_{zzz},\dots)=0
\end{equation}
(\ref{se1}) is solved by means of the ansatz
\begin{equation}\label{se2}
u(z) = \sum \limits_{k=0}^N A_k \left(\frac{\psi_k}{\psi} \right)^k
\end{equation}
where $A_k$ are constants and $A_N \ne 0$. The
function $\Psi$ is a solution of some ordinary differential equation of lesser order than (\ref{se1})
(called simplest equation) and solutions of these simplest equations are known. One uses the finite series
(\ref{se2}) in order to represent the solution $u$ through the solution of the simplest equation. In order to do this one has to determine the value of $N$ by means of balance of power of the leading terms in the relationship which is obtained after the substitution of
(\ref{se2}) in (\ref{se1}). This relationship is polynomial of $\frac{\Psi_z}{\Psi}$ and by equating  to
$0$ of the coefficients to the powers of $\frac{\Psi_z}{\Psi}$ one obtains a system of nonlinear algebraic equations which solution leads to an exact solution of
(\ref{se1}).
\par 
Let us prove that the Modified Simple Equation Method is a specific case of SEsM. 
\par 
{\bf Proposition}\\
{\em 
The Modified Simple Equation Method is specific case of SEsM for the case when there is no transformation of the nonlinearity of the equation (Step 1 of SEsM is skipped); Function $F$ at Step 2 of SEsM has specific form -  (\ref{se2}) and just one simple equation is used.}	
\par 
{\bf Proof}\\
\begin{enumerate}
	\item  
We start from SEsM, impose restrictions on it and reduce SEsM to Modified Method of Simple Equation. 
\item  
In Step 1. of SEsM we do not transform the nonlinearity of
the solved equation (we just skip this step).
Additional restriction is that we search for solution of 
the solved equation which depends on a single coordinate 
$z$ - (\ref{se1}). 
\item  
In Step 2. of SEsM we use a possible form of the function $F$ - (\ref{se2}).
This possible form is just one of the many forms that can be used in SEsM.
\item 
In Step 3.  of SEsM, the function from (\ref{se2}) in the JEFEM is directly connected to the solution of the used simple equation which in this case is called simplest equation. We note that the use of only one simple equation is a further restriction on SEsM. 
\item 
By means of all
restrictions above we reduce SEsM to the Modified Method of Simple Equation. Thus Modified Method of Simple Equation  is specific case of SEsM.
\end{enumerate} 
\section{Trial Function Method as Specific Case of SEsM}
The Trial Function Method is as follows \cite{lx1},
\cite{xie}. One consider a nonlinear partial differential
equation
\begin{equation}\label{tf1}
N(u,u_x,u_t,u_{xx},u_{xt},u_{tt},\dots)=0,
\end{equation} 
and takes a trial function $y(x,t)$ and construct a
solution $u(y)$ of (\ref{tf1}). Then we substitute
$u(y)$ in (\ref{tf1}) and determine the parameters of the
solution.
\par
The trial function can have different form. For an example the trial function in \cite{lx1} is
\begin{equation}\label{tf2}
y = y_0 + \frac{b \exp(\beta \xi)}{[1 + \exp(a \xi)]^d}.
\end{equation}
In (\ref{tf2}) $y_0, a, b, d, \beta$ are parameters and $\xi$ is the traveling-wave coordinate.

\par 
{\bf Proposition}\\
{\em 
The Trial Function Method is specific case of SEsM for the case when there is no transformation of the nonlinearity of the equation (Step 1 of SEsM is skipped); Function $F$ at Step 2 of SEsM has specific form -  $u(y)$ where $y$ (the trial function) is the solution of the just one used simple equation.}	
\par 
{\bf Proof}\\
\begin{enumerate} 
	\item  
We start from SEsM, impose restrictions on it and reduce SEsM to the Trial Function Method. 
\item  
In Step 1. of SEsM we do not transform the nonlinearity of
the solved equation (we just skip this step).
\item  
Additional restriction is that we search for solution of 
the solved equation which depends on a single coordinate 
which can be traveling wave coordinate or other kind of coordinate. 
\item  
In Step 2. of SEsM we use a specific form of the function $F$  which is constructed by
means of trial function. In the most cases $F$ is
presented by finite power series of the trial function.
The trial function is a solution of one simple equation.
\item 
Thus  by means of the
restrictions above we reduce SEsM to the Trial Function
Method. Thus Trial Function Method  is specific case of SEsM .
\end{enumerate} 
\section{General Projective Riccati Equations Method as Specific Case of SEsM}
The general projective Riccati equations method is as follows \cite{biao}. One consider the
equation
\begin{equation}\label{bi1}
P(u,u_x,u_t,u_{xx},u_{xt},u_{tt},\dots)=0.
\end{equation}
$P$ is a function of $u$ and its derivatives.
Then, one converts (\ref{bi1}) to an ordinary differential equation by means of the travelling wave ansatz
$u(x,t)=u(\xi)$, $\xi = x-\lambda t$. The resulting ordinary differential equation is
\begin{equation}\label{bi2}
G(u, u',u'',u''', \dots) = 0.
\end{equation}
$G$ is a function of $u$ and its derivatives. The methodology has the following steps. First of one
balance of the highest derivative and of the nonlinearities in (\ref{bi2}) is made.  This is made by the
substitution
\begin{equation}\label{bi3}
u(\xi) = \varphi^m(\xi).
\end{equation}
$m$ is the balance constant. After the determination of $m$ one searches for solutions of (\ref{bi2}) from the kind
\begin{equation}\label{bi4}
u(\xi) = A_0 + \sum \limits_{i=1}^m \sigma^{i-1}[A_i \sigma(\xi) + B_i \tau(\xi)].
\end{equation}
In (\ref{bi4}) $A_i$ and $B_i$ are parameters and the functions
$\sigma(\xi)$ and $\tau(\xi)$ satisfy the differential equations
\begin{eqnarray}\label{bi5}
\frac{d \sigma}{d \xi} &=& \epsilon \sigma \tau; \nonumber \\
\frac{d \tau}{d \xi} &=& R + \epsilon \tau^2 - \mu \sigma .
\end{eqnarray}
Above $\epsilon=\pm 1$. $R \ne 0$ and $\mu \ne 0$ are parameters.
For the case $R = \mu = 0$ the solution is searched in the form
\begin{equation}\label{bi6}
u(\xi) = \sum \limits_{i=0}^m A_i \tau^i(\xi).
\end{equation}
\par 
{\bf Proposition}\\
{\em 
The General Projective Riccati Equations Method is specific case of SEsM for the case when there is no transformation of the nonlinearity of the equation (Step 1 of SEsM is skipped); Function $F$ at Step 2 of SEsM has specific form -  (\ref{bi3}) or (\ref{bi4}) and the simple equation is
\begin{equation}\label{see}
\frac{1}{\epsilon} \frac{d^2}{d\xi^2}(\ln \sigma) = R + \frac{1}{\epsilon} \left( 
\frac{d \ln(\sigma)}{d \xi}\right)^2 - \mu \sigma
\end{equation}
}	
\par 
{\bf Proof}\\
We start from SEsM, impose restrictions on it and reduce SEsM to the General Projective Riccati Equations
Method. At Step 1. of SEsM we do not transform the nonlinearity of
the solved equation (we just skip this step), i.e., we consider specific
case of SEsM without transformation of nonlinearity of the solved equation.
Additional restriction is that we search for solution of 
the solved equation which depends on a single coordinate 
which can be traveling wave coordinate or other kind of coordinate. At Step 2. of SEsM we use a specific form of the function $F$  which is (\ref{bi3}) for the case $R \ne 0, \mu \ne 0$ and (\ref{bi4})
for the case $R = \mu =0$. The functions $\sigma$ and $\tau$ can be determined from (\ref{see}).
(\ref{see}) is obtained as follows. From first from the equations in (\ref{bi5}) one obtains
\begin{equation}\label{bi7}
\tau = \frac{1}{\epsilon} \frac{d(\ln \sigma)}{d \xi}
\end{equation}
The substitution of (\ref{bi7}) in the second of the equations from (\ref{bi5}) leads to (\ref{see}).
Then, we have one simple equation: (\ref{see}). We use this simple equation and $\tau$ can be
determined from $\sigma$ from (\ref{bi7}). On the basis of (\ref{see}) ,(\ref{bi4}) or (\ref{bi6})
one tries to reduce the solved equation to a system of nonlinear algebraic equations (Step 6 of SEsM).
If this is successful one may obtain an exact solution of the solved equation (Step 7 of SEsM).
Thus, the General Projective Riccati Equations Method  is specific case of SEsM.

\section{First Integral Method as Specific Case of  SEsM}
The First Integral Method for obtaining exact solutions 
of nonlinear partial differential equations is as 
follows \cite{f02}. One wants to obtain exact solution of the nonlinear partial differential equation
\begin{equation}\label{fe1}
P(u,u_x,u_t,u_{xx},u_{xt},u_{tt},\dots)=0,
\end{equation} 
One converts (\ref{fe1}) to ordinary differential equation by the traveling wave ansatz $u(x,t)=U(z) = u(kx-\omega t)$. Then one introduces $X=U$ and $Y=U_z$ and writes (\ref{fe1}) as system of equations
\begin{equation}\label{fe2}
Y=X_z
\end{equation}
\begin{equation}\label{fe3}
Y_z=F(X,Y)
\end{equation}
The solution is obtained by the assumption that the 
derivative of the
relationship $Q(X,Y) = \sum \limits_{i=0}^m a_i(X) Y^i$
can be represented as
\begin{equation}\label{fe4}
\frac{dQ}{dz} = [g(X) + h(X)Y] \sum \limits_{i=0}^m a_i(X)Y^i
\end{equation}
which together with (\ref{fe3}) allow computation of the
solution.
\par 
{\bf Proposition}\\
{\em 
The First Integral Method is specific case of SEsM for the case when equations of the kind 
\begin{equation}\label{fe5}
X_{zz} = F(X,X_z)
\end{equation}
are considered,
there is no transformation of the nonlinearity of the equation (Step 1 of SEsM is skipped); single simplest equation is used and this simplest equation is determined by the condition (\ref{fe4})	}
\par 
{\bf Proof}\\
\begin{enumerate}
	\item   
We note that the First Integral Method can be applied
to the restricted class of equations (\ref{fe5}). This
restricted class is obtained from(\ref{fe3}) by 
substitution of (\ref{fe2}) there.
\item  
We start from SEsM, impose restrictions on it and reduce SEsM to the Trial Function Method. 
\item 
In Step 1. of SEsM we do not transform the nonlinearity of
the solved equation (we just skip this step).
\item 
Additional restriction is that we search for solution of 
the solved equation which depends on a single coordinate 
which can be traveling wave coordinate or other kind of coordinate. (\ref{fe4}) imposes further restriction on $X$ and plays the role of implicit simple equation which together with (\ref{fe5}) determine the solution of (\ref{fe1}). 
\item  
In this process one has to use polynomial
form of $a_i(X)$ and to determine the coefficients of these polynomials similar to the steps of SEsM.
\item  
This First Integral Method is specific case of SEsM for
obtaining solutions for the limited c lass of equations
(\ref{fe3}) under the assumption that (\ref{fe4}) holds .
\end{enumerate} 
\section{Concluding  Remarks}
We discuss in this article the methodology of SEsM (the Simple Equations Method) as well as  the  relations of this methodology to  several other methods for obtaining exact solutions of nonlinear differential equations.
\begin{itemize} 
	\item  
We show that numerous methods are specific cases of SEsM. These methods use different forms of the (in the most cases single)
simple equation. 
\item 
In addition almost all of these methods search for solutions which are constructed as power series
of the solution of the considered simple equation. Usually the corresponding method takes the name of the used
simple equation. 
\item  
SEsM does not prescribe the form and the number of the used simplest equations. 
\item  
SEsM does not fix the 
relationship among the solution of the solved equation and the solutions of the used simple equation(s). 
\item 
Because of this SEsM
is a general method which has numerous specific cases. We intent to continue this research in order to
find the methods for obtaining exact solutions of nonlinear differential equations which are not specific cases of SEsM.
\end{itemize}

\end{document}